\documentclass[12pt]{article}

\usepackage{cite}
\usepackage{epsf}
\usepackage{graphicx}
\usepackage{psfrag}
\usepackage[dvips]{epsfig}

\usepackage[english]{babel}

\usepackage{amssymb,indentfirst}
\usepackage{amsmath}

\usepackage{verbatim}

\makeatletter
\renewcommand \thesection {\@arabic\c@section.}
\renewcommand\thesubsection   {\thesection\@arabic\c@subsection.}
\renewcommand\thesubsubsection{\thesubsection\@arabic\c@subsubsection.}
\makeatother

\def\starup#1{\mbox{$\raise1.8ex\hbox{$*$} \kern-.7em#1$}}
\def\krup#1{\mbox{$\raise1.8ex\hbox{$+$} \kern-1.0em#1$}}
\def\linup#1{\mbox{$\raise1.9ex\hbox{---} \kern-1.0em#1$}}

\begin{document}
\title{Chiral color symmetry
\\ and possible $G'$-boson effects
\\ at the Tevatron and LHC
}

\author{M.V.~Martynov\footnote{E-mail: martmix@mail.ru}, \,
 A.D.~Smirnov\footnote{E-mail: asmirnov@univ.uniyar.ac.ru}\\
{\small Division of Theoretical Physics, Department of Physics,}\\
{\small Yaroslavl State University, Sovietskaya 14,}\\
{\small 150000 Yaroslavl, Russia.}}
\date{}
\maketitle


\begin{abstract}
\noindent
A gauge model with chiral color symmetry is considered and possible effects
of the color $G'$-boson octet predicted by this symmetry are investigated in dependence
on two free parameters, the mixing angle  $\theta_G$ and $G'$ mass $m_{G'}$.
The allowed region in the $m_{G'} - \theta_G$ plane is found from the Tevatron
data on the cross section $\sigma_{t\bar{t}}$ and forward-backward asymmetry $A_{\rm FB}^{p \bar p}$
of the $t\bar{t}$ production.
The mass limits for the $G'$-boson are shown to be stronger than those for the axigluon.
A possible effect of the $G'$-boson on the $t\bar{t}$ production at the LHC is discussed
and the mass limits providing for the $G'$-boson evidence at the LHC are estimated
in dependence~on~ $\theta_G$.

\vspace{5mm}


\end{abstract}




The search for new physics beyond the Standard Model (SM) induced by higher
symmetries (such as supersymmetry, left-right symmetry, etc.) is one of the modern
research directions in elementary particle physics.
The Large Hadron Collider (LHC) will 
allow the exploration of the existence of new physics at the TeV energy scale
with very large statistics \cite{Butterworth:2007bi}.
Top physics is a very promising place
to look for new physics effects \cite{Hill:1993hs} 
and a top factory such as the LHC is expected to be a goldmine for studying the SM
as well as beyond the SM physics \cite{Beneke:2000hk}.

There are models extending the standard color gauge group $SU_c(3)$
to the group of the chiral color symmetry
%
\begin{equation}
\label{chiral_group}
G_c=SU_L(3)\!\times \! SU_R(3) \to SU_c(3),
\end{equation}
which is assumed to be valid at high energies
and is broken to usual QCD $SU_c(3)$ at low energy scale.
Such chiral color theories~\cite{Pati:1975ze,Hall:1985wz,Frampton:1987ut,Frampton:1987dn}
in addition to the usual massless gluon $G_\mu$ predict in the simplest case of $g_L=g_R$
the existence of a new color-octet gauge boson, the axigluon $G^A_\mu$ with mass $m_{G_A}$.
The axigluon couples to quarks with an axial vector structure and with the same strong
interaction coupling strength as QCD. It has a width $\Gamma_{G_A}\approx 0.1 m_{G_A}$~\cite{Bagger:1988}.
Since it is the colored gauge particle with axial vector coupling to quarks, the axigluon should immediately
result in the increase of the hadronic cross section and in the appearance of a forward-backward asymmetry
of order $\alpha_s^2$ \cite{rodrigo-2008}.
The CDF data on the cross section of the dijet production at the Tevatron~\cite{Aaltonen:2008dn}
exclude at 95\% C.L. the axigluon mass region $260$~GeV$ < m_{G_A} < 1.250$~TeV
and the Tevatron data on asymmetry sets  the lower mass limit for the axigluon at
$m_{G_A} > 1.2, 1.4$~TeV~\cite{antunano-2007,rodrigo-2008}.

The massive color octet with arbitrary vector-- and axial-vector--quark coupling constants
has been considered phenomenologicaly  in ref.~\cite{ferrario-2008}.
But it is also interesting  to consider the color octet as the gauge boson induced by
the chiral color symmetry of a general type.

In the present paper we consider the color-octet boson
induced by the gauge chiral color symmetry~\eqref{chiral_group}
in general case of~$g_L\neq g_R$.
We calculate the possible contributions of this boson to the cross section
and to the forward-backward asymmetry of the $Q \bar{Q}$ production in $p \bar{p}$ and $pp$ collisions
in dependence on the free parameters of the model.
We compare the results with the Tevatron data on the $t \bar{t}$ production and discuss
a possible effect of this boson in the $t \bar{t}$ production at the LHC.

To reproduce the usual quark-gluon interaction of QCD the gauge coupling constants $g_L, \, g_R$
of the gauge group~\eqref{chiral_group} must satisfy the relation
\begin{eqnarray}
\frac{g_L g_R}{\sqrt{(g_L)^2+(g_R)^2}} = g_{st}
\label{eq:gLgRgst}
\end{eqnarray}
where $g_{st}$ is the strong interaction coupling constant.

The basic gauge fields $G^L_\mu$ and $G^R_\mu$ are mixed and form the usual gluon field $G_\mu$
and the field $G'_\mu$ of an additional $G'$-boson as
\begin{align}
G_\mu&= \frac{g_R G^L_\mu + g_L G^R_\mu}{\sqrt{(g_L)^2+(g_R)^2}} \equiv s_G \, G^L_\mu + c_G \, G^R_\mu,\\
G'_\mu&= \frac{g_L G^L_\mu -  g_R G^R_\mu}{\sqrt{(g_L)^2+(g_R)^2}} \equiv c_G \, G^L_\mu - s_G \, G^R_\mu,
\end{align}
where $G^{L,R}_{\mu} = G^{L,R}_{i\mu} t_i$, $G_{\mu} = G^i_{\mu} t_i$, $G'_{\mu} = G'^i_{\mu} t_i$,
 $i=1,2,...,8$, $t_i$ are the generators of $SU_c(3)$ group,
$s_G =\sin\theta_G, \, c_G =\cos\theta_G $, $\theta_G$ is $G^{L} - G^{R}$ mixing angle,
$tg\,\theta_G=g_R/g_L$.

The symmetry~\eqref{chiral_group} can be softly broken by the scalar field $\Phi_{\alpha \beta}$,
which transforms according to the $(3_L, \bar{3}_R)$ representation of the group~\eqref{chiral_group}
and has the VEV
$\langle \Phi_{\alpha \beta} \rangle =  \delta_{\alpha \beta} \, \eta /(2 \sqrt{3}) $, \,
$\alpha, \beta = 1,2,3$ are the $SU_L(3)$ and $SU_R(3)$ indices.
After such symmetry breaking the gluons are still massless and the $G'$-boson acquires the mass
\begin{equation}
m_{G'} = \frac{g_{st}}{s_G c_G} \, \frac{\eta}{\sqrt{6}} .
\label{eq:MG1}
\end{equation}

The interaction of the $G'$-boson with quarks can be written in the model independent form as
\begin{equation}
\mathcal{L}_{G'qq}=g_{st} \, \bar{q} \gamma^\mu (v + a \gamma_5) G'_\mu q
\end{equation}
where $v$ and $a$ are the phenomenological vector and axial-vector coupling constants.
The gauge symmetry~\eqref{chiral_group}  gives for $v, \, a$ the expressions
\begin{equation}
v = \frac{c_G^2-s_G^2}{2 s_G c_G} = \cot(2\theta_G), \,\,\,\,
a = \frac{1}{2 s_G c_G} = 1 / \sin(2\theta_G) .
\label{eg:va}
\end{equation}

So, in the general case of the gauge chiral color simmetry~\eqref{chiral_group}
the mass of the $G'$-boson is defined by expression~\eqref{eq:MG1}
and the vector and axial-vector coupling constants of $G'$-boson with quarks
(in contrast to phenomenological approach of ref.~\cite{ferrario-2008})
depend on one parameter $\theta_G$ which is defined by the gauge coupling constants $g_L, \, g_R$
satisfying relation~\eqref{eq:gLgRgst}.
This circumstance reduces the possible region of the parameters $v, \, a$, and allows the possibility
of studing the phenomonology of the $G'$-boson in more detail in dependence on two free parameters of
the model $m_{G'}$ and $\theta_G$.
In the particular case of $g_L=g_R$ \, $\theta_G=45^\circ$, \, $v=0$, $a=1$, the $G'$-boson coincides
with the axigluon.
In the general case of decreasing $\theta_G$, the coupling constants increase according to~\eqref{eg:va}
so that, for example, for $\theta_G= 15^\circ, \, 10^\circ$
the pertubation theory parameters take the values
$\alpha_s v^2 / \pi \approx \alpha_s a^2 / \pi  \approx 0.14, \,  0.3$ respectively.
In further considerations we restrict ourselves to the mixing angle region
$10^\circ\lesssim\theta_G\leq 45^\circ$.

The hadronic width of the $G'$-boson can be written as
\begin{eqnarray}
\Gamma_{G'} = \sum_{Q} \Gamma (G' \to Q\overline{Q})
\label{GammaG'}
\end{eqnarray}
where
\begin{eqnarray}
\Gamma (G' \to Q\overline{Q}) =
\frac{\alpha_{s}\, m_{G'}}{6}
\Bigg[ \, v^2 \left(1+\frac{2m_Q^{2}}{m_{G'}^{2}}\right)
+ a^2 \left(1-\frac{4m_Q^{2}}{m_{G'}^{2}}\right) \Bigg] \sqrt{1-\frac{4m_Q^{2}}{m_{G'}^{2}}}
\label{GammaG'QQ}
\end{eqnarray}
is the width of $G'$-boson decay into $Q\overline{Q}$-pair.
In the case of neglecting the masses of light quarks (except t-quark)
the result \eqref{GammaG'}, \eqref{GammaG'QQ} agrees with that of ref.~\cite{ferrario-2008}.

Using the coupling constants~\eqref{eg:va} from~\eqref{GammaG'},~\eqref{GammaG'QQ} we obtain
the next estimations for the relative width of $G'$-boson 
\begin{equation}
  \Gamma_{G'}/m_{G'}=0.11, \; 0.18, \; 0.41, \; 0.75, \; 1.71
\label{gammaG1}
\end{equation}
for  $ \theta_G=45^\circ, \; 30^\circ, \; 20^\circ, \; 15^\circ, \; 10^\circ $
respectively.

Since it is a strongly interacting particle, the $G'$-boson can give significant
contributions to the production of quark--antiquark pairs in $p \bar{p}$ and $pp$ collisions.
The differential partonic cross section of the process $q\bar{q} \rightarrow Q \bar{Q}$
considering the $G'$-boson and gluon contributions within the tree approximation has been calculated
(in agreement with ref.~\cite{ferrario-2008}) and can be written as
\begin{eqnarray}
\nonumber
&&\frac{ d\sigma(q\bar{q} \stackrel{\,g,\,G'}{\rightarrow} Q \bar{Q}) }{d\cos \hat{\theta}} = 
 \frac{\alpha_s^2\pi \beta}{9\hat{s}}
\bigg \lbrace f^{(+)}+\frac{2 \hat{s} (\hat{s}-m_{G'}^2)}
{(\hat{s}-m_{G'}^2)^2+m_{G'}^2 \Gamma_{G'}^2}
\Big[ \, v^2 f^{(+)} + 2 a^2 \beta c \, \Big] +
\\ \label{diffsect}
&& + \frac{\hat{s}^2} {(\hat{s}-m_{G'}^2)^2+m_{G'}^2 \Gamma_{G'}^2}
\Big[ \left( v^2 + a^2 \right)
\big( v^2 f^{(+)}+   a^2 f^{(-)} \big)
+ 8 a^2v^2 \beta c \, \Big]
\bigg \rbrace,
\end{eqnarray}
where $f^{(\pm)}=(1+\beta^2 c^2\pm 4m_Q^2/\hat{s})$, $c = \cos \hat{\theta}$,
$\hat{\theta}$ is the  scattering angle of $Q$-quark in the parton center of mass frame,
$\hat{s}$ is the invariant mass of $Q \bar{Q}$ system,
$\beta = \sqrt{1-4m_Q^2/\hat{s}}$.

Integration of~\eqref{diffsect} over the angle gives the corresponding total cross section in the form
\begin{eqnarray}
\nonumber
\sigma(q\bar{q} \stackrel{\,g,\,G'}{\rightarrow} Q \bar{Q}) &=& \frac{4\pi \alpha_s^2\beta}{27\hat{s}}
\bigg \lbrace
3-\beta^2 - \frac{2\hat{s}m_{G'}^2v^2(3-\beta^2)}{(\hat{s}-m_{G'}^2)^2+\Gamma_{G'}^2 m_{G'}^2}+\\
&+&\frac{\hat{s}^2 \big[ \, (v^4+2v^2)(3-\beta^2) + v^2 a^2 (3+\beta^2) + 2a^4\beta^2 \, \big]}
{(\hat{s}-m_{G'}^2)^2 + \Gamma_{G'}^2 m_{G'}^2}
\bigg \rbrace.
 \label{sect}
\end{eqnarray}

In tree approximation, the $G'$-boson does not contribute  to the $g g  \rightarrow Q \bar{Q}$ process
of $Q \bar{Q}$ production in gluon fusion. The differential and total SM partonic cross sections
of this process are well known and have the form
\begin{equation}
\frac{d\sigma(gg\rightarrow Q \bar{Q})}{d\cos \hat{\theta}} =
\alpha_s^2 \: \frac{\pi \beta}{6 \hat{s}}
\left(\frac{1}{1-\beta^2c^2}-\frac{9}{16}\right)
\left(1 + \beta^2 c^2 +2(1-\beta^2)-\frac{2 (1-\beta^2)^2}{1-\beta^2 c^2}\right),
\label{difcsggQQ}
\end{equation}
\begin{equation}
\sigma(gg\rightarrow Q \bar{Q}) = \frac{\pi  \alpha_s^2 }{48 \hat{s}}
\left[
\left(\beta ^4-18 \beta ^2+33\right) \log \left(\frac{1+\beta }{1-\beta }\right)+
\beta  \left( 31 \beta ^2-59 \right)
\right].
\label{totcsggQQ}
\end{equation}
%

Taking into account the parton densities and the values of the $\beta$ parameter in~\eqref{sect}
one can see from~\eqref{sect} that the contribution of the $G'$-boson to $Q \bar{Q}$ production
is most significant for  $t\bar{t}$ production.
The $t\bar{t}$ production is well studied at the Tevatron and
the recent CDF result for the $t\bar{t}$ production cross section is~\cite{Lister:2008it}
\begin{equation}
\sigma_{t\bar{t}}  = 7.0 \pm 0.3 (stat) \pm 0.4 (syst) \pm 0.4 (lumi) pb .
\label{expcspptt}
\end{equation}

We have calculated the cross section $\sigma(p \bar{p} \rightarrow t \bar{t})$
of $t\bar{t}$-pair production in $p\bar{p}$-collisions at the Tevatron energy
using the parton cross sections~\eqref{sect},~\eqref{totcsggQQ} and
the parton densities AL'03~\cite{alekhin} (NLO, fixed-flavor-number,  $Q^2=m_t^2$)
with the appropriate K-factor $K=1.24$~\cite{campbell-2007-70}.
%
The allowed region in $m_{G'} - \theta_G$ plane
(the unshaded region) which is compatible with data~\eqref{expcspptt} within $2 \sigma$
is shown in Fig.1,
the $1 \sigma$ region is marked by the dashed line.
From Fig.1  
and by  comparing the calculated cross section $\sigma(p \bar{p} \rightarrow t \bar{t})$
with CDF result~\eqref{expcspptt}
we find that the $G'$-boson with masses
\begin{equation}
m_{G'}[TeV]>0.91 (0.94), \, \; 1.03 (1.05), \,  \; 1.20 (1.23), \,  \; 1.39 (1.44), \,  \;
1.73 (1.82)
\label{eq:mG1limcs}
\end{equation}
is compatible with data~\eqref{expcspptt} within $2 \sigma (1 \sigma)$
for $ \theta_G=45^\circ, \; 30^\circ, \; 20^\circ, \; 15^\circ, \; 10^\circ$ respectively.
The first value in~\eqref{eq:mG1limcs} coincides with the known mass limit
for the axigluon~\cite{choudhury-2007} whereas the next ones are the mass limits
for the $G'$-boson in dependence on the mixing angle $\theta_G$.

The $G'$ boson can generate, at tree-level, a forward-backward asymmetry through the interference
of $q\bar{q} \stackrel{\,G'}{\rightarrow} t\bar{t}$ and $q\bar{q} \stackrel{\,g}{\rightarrow} t\bar{t}$
amplitudes~\cite{antunano-2007,Sehgal:1987wi,choudhury-2007}.
From~\eqref{diffsect} we find that the $G'$ boson induces a forward-backward difference
in the $q\bar{q}\to Q\bar{Q}$ cross section of the form
\begin{eqnarray}
\Delta_{FB}(q\bar{q}\to Q\bar{Q})&=& \sigma(q\bar{q}\rightarrow Q \bar{Q}, \, \cos \theta > 0)-
\sigma(q\bar{q}\rightarrow Q \bar{Q}, \, \cos \theta < 0)=
\nonumber
\\
&=&
\frac{4\alpha_s^2\pi \beta^2 a^2}{9}
\Bigg( \frac{ \hat{s}-m_{G'}^2 +2 v^2 \hat{s}}
{(\hat{s}-m_{G'}^2)^2+m_{G'}^2 \Gamma_{G'}^2}
\Bigg).
\label{eq:deltaFBqq}
\end{eqnarray}

As seen from~\eqref{eq:deltaFBqq} in dependence on the values  $\hat{s}$, $m_{G'}^2$ and $v$,
the $G'$ boson can give a contribution to the forward-backward asymmetry $A_{\rm FB}^{p \bar p}$
in $p \bar p$ collisions, which can take  positive values as well as  negative ones.
Concerning the gluon-gluon fusion, one can see from~\eqref{difcsggQQ} that this process
does not contribute to forward-backward asymmetry of the tree level.

The forward-backward asymmetry of top quarks has been measured at the Tevatron.
The latest CDF analysis \cite{aaltonen-2008} based on 1.9 $fb^{-1}$
integrated luminosity gives
\begin{equation}
    A_{\rm FB}^{p \bar p} = \frac{N_t (\cos \theta >0)-N_t (\cos \theta <0)}
{N_t(\cos \theta >0)+N_t(\cos \theta <0)}=0.17 \pm 0.07~(\rm{stat}) \pm 0.04~(\rm{sys}).
\label{AFBpptt}
\end{equation}

Using~\eqref{diffsect} (or~\eqref{eq:deltaFBqq}) and the parton densities one can
calculate the forward-backward asymmetry $A_{\rm FB}^{p \bar p}$ in dependence on
$m_{G'}$ and $\theta_G$.
The allowed region in the $m_{G'} - \theta_G$ plane
(the undashed region), which is compatible with data~\eqref{AFBpptt} within $2 \sigma$
is shown in Fig.1.
The border of the allowed $1 \sigma$ region is shown by the dashed line.
One can see that the $1 \sigma$ region allowed by $A_{\rm FB}$ data~\eqref{AFBpptt} is excluded
by the cross section data~\eqref{expcspptt}. Nevertheless, there is a region
in the $m_{G'} - \theta_G$ plane that is compatible with the data~\eqref{AFBpptt}
and~\eqref{expcspptt} simultaneusly within $2 \sigma$ (the clean region).
Comparing the calculated $A_{\rm FB}^{p \bar p}$ asymmetry
with the data~\eqref{AFBpptt} and accounting for the mass limits~\eqref{eq:mG1limcs}
from Fig.1  
we find that the $G'$-boson with masses
\begin{equation}
m_{G'}>1.44\, TeV, \; 1.56 \, TeV, \; 1.76 \, TeV
\label{eq:mG1limcsAFB1}
\end{equation}
for $ \theta_G=45^\circ, \; 30^\circ, \; 20^\circ $
as well as with masses
\begin{equation}
m_{G'} = 1.20-1.32 \, TeV, \; > 1.39 \, TeV, \; > 1.73 \, TeV
\label{eq:mG1limcsAFB2}
\end{equation}
for $ \theta_G=20^\circ, \; 15^\circ, \; 10^\circ $
is compatible with data~\eqref{AFBpptt} and~\eqref{expcspptt} simultaneusly within $2 \sigma$.
The first value in~\eqref{eq:mG1limcsAFB1} is close to the known mass limit
for the axigluon~\cite{antunano-2007} resulting from the $A_{\rm FB}^{p \bar p}$ data
whereas the other values in~\eqref{eq:mG1limcsAFB1}
and~\eqref{eq:mG1limcsAFB2} are the new mass limits
for the $G'$-boson resulting from the data~\eqref{AFBpptt}
and~\eqref{expcspptt} simultaneusly  in dependence on the mixing angle $\theta_G$.

In $pp$ collisions at the LHC the $q\bar{q}$ fluxes are essentially smaller
than the $gg$ fluxes, and $t\bar{t}$ production is dominated by the contribution from the $gg$ initial state.
However, by  increasing the $t\bar{t}$ invariant masses this dominance becomes to be less significant
and it is reasonable to search for the $G'$-boson through its effect on the $t\bar{t}$ invariant
mass distribution. The large number of  top pairs expected to be produced at the LHC
(8 million pairs for 10~$fb^{-1}$ integrated luminosity) makes a study of such a
differential distribution meaningful.

Using the parton cross sections~\eqref{sect},~\eqref{totcsggQQ} and integrating them with
the parton densities~\cite{alekhin} over the final $t$ quark rapidity $y$, we have obtained
the $t\bar{t}$ invariant mass distribution
$d\sigma_s(pp  \to  t\bar{t})/dM_{t\bar{t}}$,
which can be expected at the LHC when taking into account the $G'$-boson
contribution.
The background distribution $d\sigma_b(pp  \to  t\bar{t})/dM_{t\bar{t}}$
is obtained analogously but by neglecting the $G'$-boson contribution in~\eqref{sect}.
The former distribution exceeds the latter one and has the peak from the $G'$-boson
defined by the mass $m_{G'}$ and width~\eqref{gammaG1} of the $G'$-boson.

To distinguish the signal and background events we use the significance estimator \cite{Bartsch:824351}
\begin{equation}\label{Significance}
    \mathcal{S}=\sqrt{ 2  \big[ \, (N_s+N_b)\ln{\left(1+N_s/N_b \right)}-N_s \, \big]} \, ,
\end{equation}
where $N_s$ and $N_b$ are number of signal and background events in
the $t\bar{t}$ invariant  mass region $m_{G'}\pm \Delta M$.
These numbers can be calculated as
\begin{equation}
N_{s,b}= L\sigma_{s,b}(m_{G'},\Delta M), \qquad
\sigma_{s,b}(m_{G'},\Delta M)=\int_{m_{G'}-\Delta M}^{m_{G'}+\Delta M}
\frac{d\sigma_{s,b}(pp \to t\bar{t})}{dM_{t\bar{t}}} dM_{t\bar{t}},
\end{equation}
where $L$ is integrated luminosity and the integration mass region $\pm \Delta M$ is chosen
to maximize the significance estimator $\mathcal{S}$.
Below we take $\Delta M=1.28 \, \Gamma_{G'}$, which corresponds to the $3 \sigma$ width
in the case of a Gaussian distribution.

We have calculated and analysed the integrated luminosity which is necessary
for the evidence of $G'$-boson at the LHC.
The integrated luminosity at $3\sigma$ significance ($\mathcal{S}=3$)
in dependence on $G'$ mass for different $\theta_G$ is shown in Fig.2.
From this figure
we find that for $ \theta_G=45^\circ, \; 30^\circ, \; 20^\circ, \; 15^\circ $
the $G'$-boson with masses
\begin{equation}
m_{G'} < 6.5 \, TeV, \; 7.0 \, TeV, \; 7.9 \, TeV, \; 9.8 \, TeV
\label{eq:mG1LHC}
\end{equation}
can be evident in $t\bar{t}$ events at the LHC at integrated luminosity $L=10\,fb^{-1}$
with $3\sigma$ significance and expected numbers of signal (background) events
$N_s(N_b)=3.2(0.4), \,$ $ 3.1(0.3), \,$ $ 3.9(0.7), \,$ $ 7.0(3.6)$ respectively.
The first value in~\eqref{eq:mG1LHC} corresponds to the case of the axigluon.

It should be noted that the chiral extension~\eqref{chiral_group} of the usual $SU_c(3)$ color symmetry
and its unification with the electroweak symmetry by the group $G_c \!\times \! SU_L(2) \! \times \! U(1) $
naturally extends the Higgs sector.
For giving the masses to the up and down quarks and to the leptons
one needs two scalar doublets $(\Phi^{(1,2)}_{a})_{\alpha \beta}$
with the SM hypercharges $Y^{SM}=\mp 1 $ and VEVs
$\langle (\Phi^{(b)}_{a})_{\alpha \beta} \rangle =
\delta_{\alpha \beta} \,\delta_{ab} \,\eta_{b}/(2\sqrt{3})$
and a colorless doublet $\Phi^{(3)}_{a}$ with VEV
$\langle \Phi^{(3)}_{a} \rangle = \delta_{a2} \,\eta_{3}/\sqrt{2}$,
here $a=1,2$ is the $SU_L(2)$ index and
$\sqrt{\eta_{1}^2 + \eta_{2}^2  + \eta_{3}^2} = \eta_{SM} \approx 250 \, GeV $ is the SM VEV.
The doublets $(\Phi^{(1,2)}_{a})_{\alpha \beta}$ break the chiral symmetry~\eqref{chiral_group}
but their VEVs $\eta_{1}, \eta_{2}$ are insufficient to give
the necessary masses~\eqref{eq:mG1limcsAFB1},~\eqref{eq:mG1limcsAFB2} to the $G'$-boson
and by this reason one needs an additional scalar field $\Phi^{(0)}_{\alpha \beta}$ which
does not interact with fermions and has the VEV 
$\langle \Phi^{(0)}_{\alpha \beta} \rangle =  \delta_{\alpha \beta} \, \eta_0 /(2 \sqrt{3}) $. 
In this case the $G'$ mass can be given by
expression~\eqref{eq:MG1} with $\eta = \sqrt{\eta_{1}^2 + \eta_{2}^2  + \eta_{0}^2}$
and from~\eqref{eq:MG1},~\eqref{eq:mG1limcsAFB2} we find that the VEV of the chiral color symmetry
breaking $\eta_0$ can be relatively small, $\eta_0  \gtrsim 800 \, GeV$.

Because of the decomposition $(3_L, \bar{3}_R) = 1_{SU_c(3)} + 8_{SU_c(3)}$
the multiplets $(\Phi^{(1,2)}_{a})_{\alpha \beta}$, $\Phi^{(0)}_{\alpha \beta}$
after the chiral color symmetry breaking give rise to the $SU_c(3)$ octets
$(\Phi^{(1,2;8)}_{a})_{\alpha \beta} = \Phi^{(1,2)}_{ia} (t_i)_{\alpha \beta} $, 
$\Phi^{(0;8)}_{\alpha \beta} = \Phi^{(0)}_{i} (t_i)_{\alpha \beta} $
and to the color singlets
$(\Phi^{(1,2;0)}_{a})_{\alpha \beta} = \Phi^{(1,2)}_{0a} \, \delta_{\alpha \beta} / \sqrt{6} $, 
$\Phi^{(0;0)}_{\alpha \beta} = \Phi^{(0)}_{0} \, \delta_{\alpha \beta} / \sqrt{6} $.
The colorless doublets $\Phi^{(1,2)}_{0a}$, $\Phi^{(3)}_{a}$ form the SM Higgs doublet $\Phi^{(SM)}_{a}$
with the SM VEV $\eta_{SM}$ and
two additional doublets $\Phi'_{a}$, $\Phi''_{a}$.
So, the chiral color symmetry in addition to the new gauge $G'$-boson predicts the new scalar fields:
the colorless doublets $\Phi'_{a}$, $\Phi''_{a}$, two doublets of color
octets $\Phi^{(1,2)}_{ia}$,
the color octet $\Phi^{(0)}_{i}$ and the colorless $SU_L(2)$ singlet 
$\Phi^{(0)}_{0} = (\eta_{0} + \chi^{(0)}_0 + i \, \omega^{(0)}_0)/\sqrt{2} $
with the VEV $\eta_0$.
It should be noted that scalar octets of the different origin are predicted also
in a number of models~\cite{popov-2005-20, PPSmPhAN2007, MW, GrWise, Perez, Perez2, Choi}.
Since they are colored particles, the color scalar octets due to their interactions with gluons
can be produced in $pp$ collisions and the phenomenology of such particles at the LHC
is under active discussion
now~\cite{MW, GrWise, Perez, Perez2, Choi, Gerbush, Zerwekh, MartSmMPLA23, Dobrescu, Idilbi}.
As concerns the field $\Phi^{(0)}_{0}$ its real part $\chi^{(0)}_0$ after the chiral 
color symmetry breaking acquires the mass of order of $\eta_{0}$ whereas the imaginary 
part $\omega^{(0)}_0$ is still massless in the tree approximation and in the unitary gauge 
is not ruled out by a gauge transformation.    
The features of these new fields will be discussed in more details elsewhere.

In conclusion, we summarize the results found in this work.

The gauge model with the chiral color symmetry of quarks as a possible extension of the Standard Model
is considered, and possible effects of the color $G'$-boson octet predicted by this symmetry at
the Tevatron and LHC energies are investigated. The hadronic width of the $G'$-boson and the $G'$-boson
contributions to the cross section $\sigma_{t\bar{t}}$ and to the forward-backward asymmetry
$A_{\rm FB}^{p \bar p}$ of $t\bar{t}$ production at the Tevatron
are calculated and analysed in dependence on two free parameters of the model, the mixing angle $\theta_G$
and $G'$ mass $m_{G'}$. The allowed region in the $m_{G'} - \theta_G$ plane is found from the Tevatron
data on $\sigma_{t\bar{t}}$ and $A_{\rm FB}^{p \bar p}$. The mass limits for the $G'$-boson are shown
to be stronger than those for the axigluon due to the specific dependence of the $G'$-boson
coupling constants on $\theta_G$. A possible effect of the $G'$-boson on the $t\bar{t}$-pair
production at the LHC is discussed and the mass limits providing for the $G'$-boson evidence
at the LHC with $3\sigma$ significance at the integrated luminosity $L=10\,fb^{-1}$ are estimated
in dependence~on~$\theta_G$.


\newpage


\newpage

{\Large\bf Figure captions}

\bigskip

\begin{quotation}
\noindent
Fig. 1. The $m_{G'} - \theta_G$ regions compatible within $2 \sigma$ with CDF data
on $\sigma_{t\bar{t}}$ (the unshaded region)
and on $A_{\rm FB}^{p \bar p}$ (the undashed region).
The dashed lines denote the corresponding $1 \sigma$ regions.
\end{quotation}

\begin{quotation}
\noindent
Fig. 2. The integrated luminosity $L$ needed for $3\sigma$ evidence of $G'$-boson at the LHC
in dependence on the $G'$ mass for different $\theta_G$.
The horizontal dashed line denotes $L=10\,fb^{-1}$.
\end{quotation}


\newpage
\begin{figure}[htb]
\vspace*{0.5cm}
 \centerline{
\epsfxsize=0.7\textwidth
\epsffile{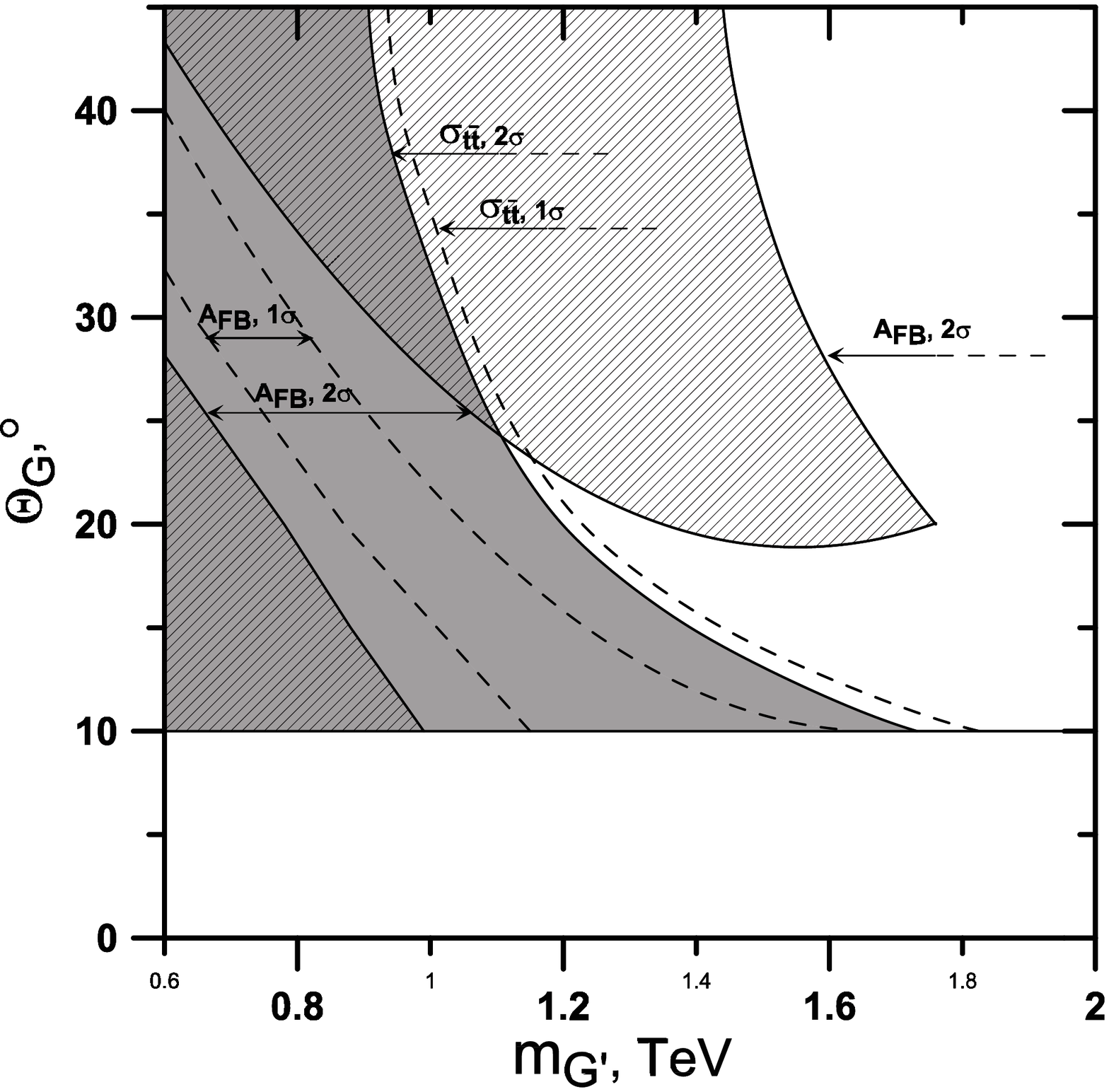}}
\vspace*{1mm}
\caption{}
\label{constrTevatron}
\end{figure}

\vspace*{5cm}
\vfill \centerline{M.V. Martynov, A.D.~Smirnov, Modern Physics Letters A}
\centerline{Fig. 1}

\newpage
\begin{figure}[htb]
\vspace*{0.5cm}
 \centerline{
\epsfxsize=0.7\textwidth
\epsffile{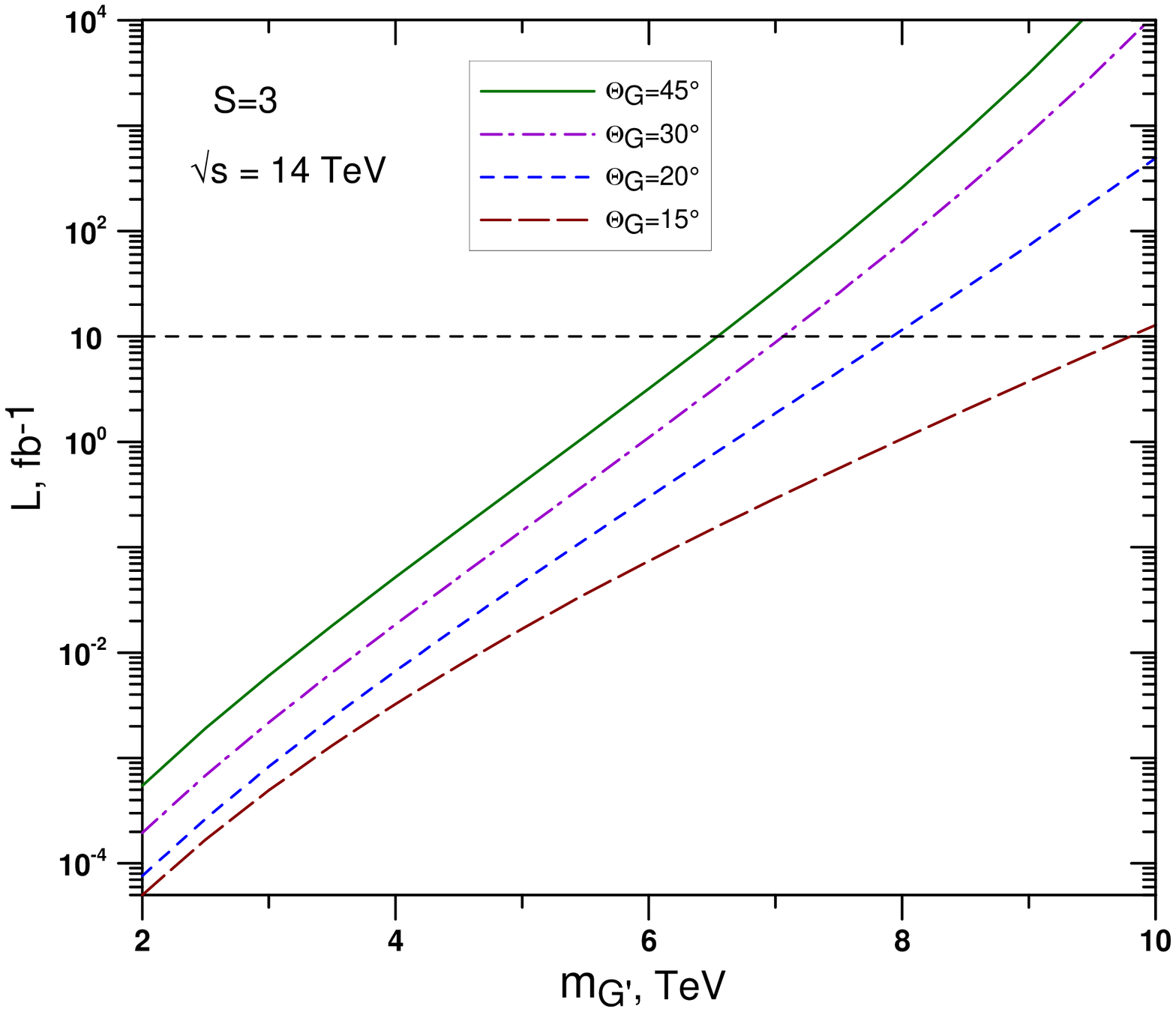}}
\vspace*{1mm}
\caption{}
\label{LHCsign}
\end{figure}

\vspace*{5cm}
\vfill \centerline{M.V. Martynov, A.D.~Smirnov, Modern Physics Letters A}
\centerline{Fig. 2}


\end{document}